# High-gradient High-charge CW Superconducting RF gun with CsK$_2$Sb photocathode

Igor Pinayev[2], Vladimir N. Litvinenko[1,2]*, Joseph Tuozzolo[2], Jean Clifford Brutus[2], Sergey Belomestnykh[2,1], Chase Boulware[3], Charles Folz[2], David Gassner[2], Terry Grimm[3], Yue Hao[2,1], James Jamilkowski[2], Yichao Jing[2,1], Dmitry Kayran[2,1], George Mahler[2], Michael Mapes[2], Toby Miller[2], Geetha Narayan[2], Brian Sheehy[2], Triveni Rao[2], John Skaritka[2], Kevin Smith[2], Louis Snydstrup[2], Yatming Than[2], Erdong Wang[2], Gang Wang[2,1], Binping Xiao[2], Tianmu Xin[1], Alexander Zaltsman[2], Z. Altinbas[2], Ilan Ben-Zvi[2,1], Anthony Curcio[2], Anthony Di Lieto[2], Wuzheng Meng[2], Michiko Minty[2], Paul Orfin[2], Jonathan Reich[2], Thomas Roser[2], Loralie A. Smart[2], Victor Soria[2], Charles Theisen[2], Wencan Xu[2], Yuan H. Wu[1], Zhi Zhao[2]

[1] Department of Physics, Stony Brook University, Stony Brook, NY, USA
[2] Brookhaven National Laboratory, Upton, NY, USA
[3] Niowave Inc., 1012 N. Walnut St., Lansing, MI, USA

PACs:85.25.-j, 07.77.Ka, 29.25.Bx, 42.72.-g, 29.20.-c, 29.20.Ej

High-gradient CW photo-injectors operating at high accelerating gradients promise to revolutionize many sciences and applications. They can establish the basis for super-bright monochromatic X-ray free-electron lasers, super-bright hadron beams, nuclear-waste transmutation or a new generation of microchip production. In this letter we report on our operation of a superconducting RF electron gun with a record-high accelerating gradient at the CsK$_2$Sb photocathode (i.e. ~ 20 MV/m) generating a record-high bunch charge (i.e., 3 nC). We briefly describe the system and then detail our experimental results. This achievement opens new era in generating high-power electron beams with a very high brightness.

Superconducting radio-frequency (SRF) electron guns are frequently considered to be the favorite pathway for generating the high-quality, high-current beams needed for future high-power energy-recovery linacs. SRF guns can find unique scientific and industrial applications, such as driving high-power X-ray and EUV CW FELs [1-10], intense γ-ray sources [11-14], coolers for hadron beams [15-18], and electron-hadron colliders [19-21]. The quality of the generated electron beam – both its intensity and brightness – is extremely important for many of these applications. The beam's quality frequently is described by its brightness, i.e., the number of generated electrons divided by the bunch's emittance (defined as the phase space-volume occupied by electrons).

In this letter, we report the record performance of our SRF gun (Fig. 1) that was built for the Coherent electron Cooling (CeC) experiment at RHIC [16] that generated a 1.7-MeV CW electron-beam with a 3 nC bunch charge. This gun demonstrated CW operation with 18 MV/m accelerating field at the CsK$_2$Sb cathode at the time of the electron's photo-emission. Table 1 summarizes the key results from CeC gun and compares them with previously commissioned SRF guns.

* Corresponding author, vladimir.litvinenko@stonybrook.edu

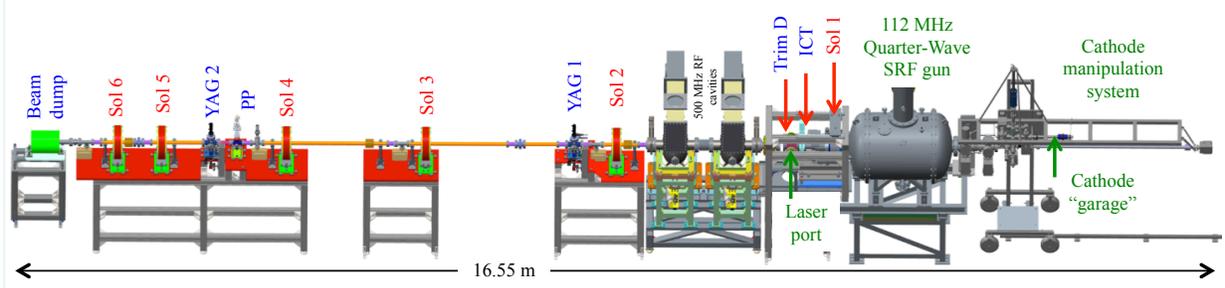

Figure 1. Layout of the CeC SRF gun beamline used in this experiment: it extends for 16.55 meters from the cathode manipulator to the beam dump. The main elements are the following: The SRF gun system, the 532 nm laser, the ICT - integrating current transformer to measure the bunch charge, the Sol1-Sol6 - focusing solenoids, the Trim D - trim dipole used for energy measurements, the YAG 1-2 - YAG screens with 1.3 MP GigE cameras for measuring the beam's profiles. PP is a multi-slit mask for emittance measurement.

It is a well-known experimental fact that both the 3D- and the projected emittances[1] grow with the increase of the bunch charge. In a photo-injector, the main method of increasing the attainable bunch charge and its brightness is via increasing the accelerating field, $\mathbf{E}_o$.

First, there is a limitation on the maximum density of the surface charge that can be extracted from a photocathode. Electron emission is affected strongly by the space-charge field. The electric field produced by a short electron bunch can be estimated from the Gauss's theorem. It limits the electron emission to the surface charge density, $\sigma$ [22] by

$$\sigma = \frac{\mathbf{E}_{cath}}{4\pi} \qquad (1)$$

where $\mathbf{E}_{cath}$ is the electric field at the cathode at the moment of emission (we note that we are using Gaussian units in this letter). Thus, the bunch charge can be increased either by increasing the beam's transverse size (and inevitably increasing its transverse emittance), or by increasing $\mathbf{E}_{cath}$.

Second, a higher accelerating gradient is required for generating high-intensity and high-quality electron beams by accelerating them as quickly to relativistic energies. While there are various contributions to the transverse emittance for a beam generated in the photo-injector, the most fundamental limitation is imposed by nonlinear space-charge forces that decrease rapidly with increasing particle's energy[2] [22]. According to the theory of emittance growth in the photo-injectors (for example [23-28]), the value of the accelerating field at the surface of the photocathode at the moment of the emission plays the most critical role. As summarized in [29-31], the overall dependence of the beam's emittance from a photo-injector can be expressed as

$$\varepsilon_n \propto \sqrt{q \frac{E_{MTE}}{\mathbf{E}_{cath}}} \qquad (2)$$

where $q$ is the bunch charge, and $E_{MTE}$ is the mean transverse-energy of the photoelectrons at the cathode. The latter depends on the photocathode composition and the laser's wavelength, which are outside of scope of our letter.

---

[1] In accelerator literature, normalized transverse emittance is defined as physical emittance divided by $mc$ ($m$ is the electron's mass, and $c$ is the speed of the light): $\varepsilon_n = \frac{1}{mc} \iint dx\, dP$. 3D normalized emittance is defined as $\varepsilon_{3Dn} = \frac{1}{(mc)^3} \iiint dx\, dP_x dy dP_y dEdt$, where $x$ and $y$ are the transverse coordinates of the particles, and $P_{x,y}$ are corresponding Canonical impulses, $t$ is the arrival time of the particle, and $E$ is it's energy.

[2] It is well known that space-charge forces are falling $\sim 1/\gamma^2$, where $\gamma = E/mc^2$ is a relativistic factor of the beam.

Currently, there are three main types of the photo-injectors: Electrostatic (DC), pulsed RF, and CW RF-electron guns. The DC guns, while generating record-performance CW beams [29], are limited in their maximum achievable accelerating gradients of 5 MeV/m. They are also limited by the (kinetic) energy of generated beam to below 0.5 MeV. In contrast, the pulsed room-temperature (normal conducting, NC) RF guns can reach accelerating gradients at the 100 MeV/m scale, but operate at relatively low repetition rate and are incapable of CW operation.

CW RF guns can be based on room temperature NC [32-33] or SRF technology [34-42]. While NC RF guns have recently demonstrated both high accelerating gradients (18 MV/m at the moment of beam emission) and a relatively large (0.3 nC) bunch charge [33], they have natural out-gassing originating from high power (~100 kW) dissipation in the cavity's walls. Accordingly, they have to operate with relatively robust photo-cathode materials (such as $Cs_2Te$), and to use UV laser light to generate photoelectrons. The latter results in a significantly higher $E_{MTE}$ and beam emittance, compared with that generated by visible laser-light from $CsK_2Sb$ photo-cathodes [43]. In contrast, with 100 kW power dissipating in a high-gradient NC RF gun, the power dissipation in an SRF gun operating at the same accelerating gradient is reduced to only a few watts. In addition, it liquid He (either 2K or 4K) cooled walls works as a very powerful cryogenic vacuum pump (absorber), therefore providing an outstanding vacuum environment. The great potential of SRF guns was recognized as early as 1988 [44] and several successful experiments have been carried out since 2002 [34-38].

Table 1. Main experimental results for five operational CW SRF photo-injectors

| Parameter | CeC | HZDR [34] | HZB [42] | NPS [37] | UW [38] |
|---|---|---|---|---|---|
| RF frequency, MHz | 112 | 1300 | 1300 | 500 | 200 |
| Cavity type | QW | Elliptical | Elliptical | QW | QW |
| Number of cells | 1 | 3.5 | 1.4 | 1 | 1 |
| LiHe temp, °K | 4 | 2 | 2 | 4 | 4 |
| Beam energy, MeV | 1.7 | 3.3 | 1.8 | 0.47 | 1.1 |
| Bunch charge, nC | 3 | 0.3 | 0.006 | 0.078 | 0.1 |
| Beam current, μA | 15 | 18 | 0.005 | <0.0001 | <0.1 |
| Dark current, nA | < 1 | 120 | - | < 20, 000 | < 0.001 |
| $\mathbf{E}_{cath}$, MV/m | 18 | 5 | 7 | 6.5 | 12 |
| Photocathode | $CsK_2Sb$ | $Cs_2Te$ | Pb | Nb | Cu |
| Laser wavelength, nm | 532 | 266 | 266 | 266 | 266 |

*QW denotes the quarter-wave cavity*

One major challenge for SRF photo-injectors is combining an RF cavity operating at cryogenic temperatures with photo-cathodes operating at room temperature[3]. SRF guns with exchangeable photocathodes require an RF-choke system, which provides effective thermal insulation and effectively

---

[3] Experiments have shown that operating $CsK_2Sb$ at the temperatures of liquid nitrogen [39] reduce quantum efficiency (QE) by about an order-of-magnitude [45]. Theory predicts that cooling it to liquid He temperatures will reduce the QE even further. In practice, replacing such a photocathode also is a long complicated procedure.

"grounds" the cathode to the nearby SRF cavity wall. An even more difficult challenge is to make the SRF cavity and the photocathode compatible. First, the cathode itself (or its insertion system) can deposit lossy particulates or materials on the walls of the cavity, and so dramatically reducing its performance. Second, the cavity can generate ion and electron back-bombardment of the cathode, and cause the cathode's rapid degradation. Practical experience with RF photo-injectors has proven both of these challenges to be very serious problems.

So far, most of the SRF guns were operating in "test mode" using metal photocathodes [37,38,42] generating a relatively low bunch charge and an average current. Metal cathodes are very robust, have very low quantum efficiency (QE), and are not suitable for generating significant beam currents. They demonstrated an accelerating gradient at the cathode from 5 to 12 MV/m.

The most advanced and well-established SRF gun, prior to our experiment, was one at HZDR (former FZD, Rossendorf), using multiple photocathode materials and operating for user experiments [34-35]. The HZDR design uses a 3.5 cell elliptical cavities operating at a frequency of 1.3 GHz. HZDR pioneered demonstration of operating SRF electron gun with an advanced photocathode material, $Cs_2Te$.

The CeC SRF gun was built by Niowave Inc. It is a part of the facility for testing coherent electron cooling [16]. It is located in the interaction region 2 (IP2) of the Relativistic Heavy Ion Collider (RHIC) and can be operated only during RHIC runs. The gun utilizes coaxial quarter-wave cavity geometry (see Fig. 2) and operates at a low frequency of 112 MHz. A half-wavelength RF-choke's design is similar to chokes previously tested in three QW SRF guns built by Niowave Inc. for Naval Postgraduate School [37] and University of Wisconsin [38]. The RF-choke incorporates a hollow copper cathode stalk allowing insertion of a cathode puck. The stalk has room temperature and it is gold-plated to reduce the heat emissivity. It uses an impedance transformer in its middle to reduce the current through the short. A pick-up antenna for measuring the RF voltage is also located inside the choke.

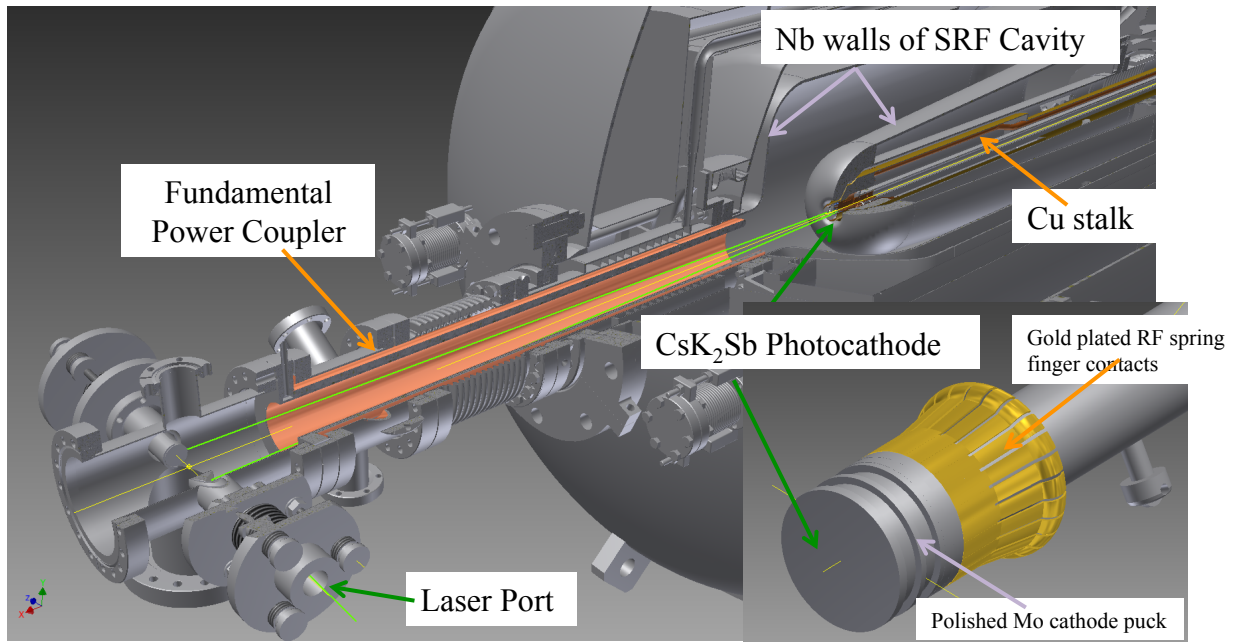

Figure 2. Details of the 112 MHz QW SRF gun. Polished Mo cathode puck is used for depositing the $CsK_2Sb$ photocathode.

The improvement over prior designs [37,38] was capability of *in-situ* cathode pucks replacements using a manipulator (Fig. 2). A small puck, shown in Fig. 2 is made of molybdenum. A $CsK_2Sb$ photoemission layer is deposited on the polished front surface of the puck. Two grooves on the side of the puck allow its inserting and extracting from the cathode deposition, the SRF gun, the load-lock and the cathode-manipulation systems. Up to three cathode pucks were stored and transported in the ultra-high vacuum (UHV) portable system (which we call a "garage") from the cathode-deposition site at Instrumentation Division (BNL) to the RHIC tunnel and back. The "garage" has built-in QE measuring system as well as a UHV vacuum gauge allowing us to monitor the state of the cathodes. Ion and NEG pumps in the garage maintain the low base pressure during cathode transit.

Our choice of RF frequency and geometry for the CeC SRF gun significantly increase the accelerating field at the moment of the beam's emission, compared with the HZDR design. The accelerating gradient realized at the photocathode at the moment of the beam's emission depends not only on the maximum electric field attainable in the RF gun, $\mathbf{E}_o$. It is also strongly depends on the emission's phase (or time $t_o = \phi_o / \omega_{rf}$) when the electrons are generated at the photocathode, $\mathbf{E}_{cath} = \mathbf{E}_o \cdot \sin\phi_o$ [22]. The optimum emission phase (time) in the electron gun is selected to maximize the beam's energy gain in the gun. It depends on the RF cavity's geometry, accelerating gradient $\mathbf{E}_o$, and the RF cavity's frequency, $f_{rf} = \omega_{rf} / 2\pi$ and is well described by a dimensionless parameter [28]:

$$\alpha = \frac{e\mathbf{E}_o}{2mc^2 k_{rf}} \quad (2)$$

where $k_{rf} = \omega_{rf} / c$ is the RF wavenumber. The value of $\alpha$ indicates the relativism of the particle exiting the first RF cavity [22,28] as well as the optimum phase of the emission: For $\alpha < 1$, the optimum emission phase is close to zero and for $\alpha \gg 1$ it is close to the crest of the accelerating field (i.e., 90°). Specifically, for the HZDR SRF gun and attainable $\mathbf{E}_o \sim$ 20 MV/m ($\alpha = 0.7$) the optimum emission phase is from 10° to 15° and $\mathbf{E}_{cath} \sim (0.2 - 0.25) \cdot \mathbf{E}_o$. In the CeC SRF gun operating that the same accelerating gradient ($\alpha = 8.34$) the optimum emission-phase is 78.5°, and there is practically no reduction in the accelerating field observed by the emitted electrons: $\mathbf{E}_{cath} \cong 0.98 \cdot \mathbf{E}_o$. This is one of the main reasons our gun operated with a four-fold higher $\mathbf{E}_{cath}$ than HZDR gun.

The operation of our SRF guns has peculiarities connected to its location inside the RHIC tunnel, and its use of liquid He provided by the RHIC's cryogenic system. During the RHIC's run, we can access the system only during maintenance periods, which are scheduled every two weeks for about 8 hours.

The gun was installed into RHIC's IR2 in the fall of 2014 and was conditioned to a 1.3 MV level of CW performance by early December 2014 with a blank (uncoated) molybdenum cathode puck. We overcame a number of multipacting zones associated with coaxial structures in the fundamental power coupler and cathode stalk using our 2 kW RF transmitter and low-level RF system [40,41,46]. Since December we experienced several mechanical and vacuum faults in the cathode exchange system[4]. Each of these failures delayed the gun commissioning by 1.5 months. Meanwhile, several very good CsK$_2$Sb gun photocathodes (with a high QE from 5% to 10%) had been manufactured and tested for survival in a new "garage". The latter allowed us "in-situ" monitoring of the cathode's QE. Significant losses in QE (for example from 5% to 3.5%) occurred only from outgassing of a Viton seal. when we opened or closed vacuum valve of the garage. Otherwise, the cathodes survived for many days without noticeable degradation of the QE: for example, QE was reduced from 3% to 2.75% inside the "garage" during 12 days while being connected to the gun's load-lock system. On May 27, 2015, we finally installed a functional CsK$_2$Sb photocathode into the gun and observed new and very significant multipacting zones. Our intense studies showed that these new zones are associated with presence of the CsK$_2$Sb on the sides of the cathode puck. Specifically, computer simulations [47] showed that the multipacting could occur between the side of the cathode plug and the stalk if the secondary emission yield coefficient exceeds 100, which is possible with such a highly emissive material as CsK$_2$Sb. Our cathode deposition system does not have a mask and the cathode material also is deposited on the puck's sides. A similar problem was observed in the HZDR's SRF gun with its $Cs_2Te$ photocathode [35], even though geometry of their RF choke has a very different geometry.

---

[4] The system was prone to serious vacuum leaks, and we are replacing it with a reliable system.

Attempts to overcome this multipacting were not successful. We decided to remove the $CsK_2Sb$ material from sides of one cathode using a previously developed technique of UV laser-cleaning [55]. This process generated bursts of the high pressure outgassing also completely extinguishing QE from the cathode's front surface. Fortunately, baking the cathode for two hours at 80 C$^o$ partially restored the QE of the cathode to the 0.8% level [48]. We used this laser-cleaned cathode for our last-chance opportunity of generating the beam from the gun during the remaining two days of the RHIC cryo-system operation in Run 15. In preparation for this event we also used helium conditioning of the SRF gun to reduce dark (cold-emission) current. This treatment improved the performance of the gun to 1.7 MV in CW mode, and 2 MV in pulsed mode. We also commissioned the low level RF controls and a green pulsed laser with maximum pulse rate of 5 kHz.

On June 24, 2015, we inserted the active photo-cathode into the gun and immediately established the 1.56 MV CW operation. With the laser on and synchronized with the RF, we scanned the phase of the laser and immediately observed photo-emitted electron bunches using an integrating current transformer (ICT) installed at the exit of the SRF gun. We proved that the bunches were generated by photo-emission by turning on and off the laser pulses. The charge of the generated electron bunches depended both on the energy and the RF phase of laser pulse.

With an initial size of the laser spot on the photocathode of ~1.5 mm FWHM (~0.7 mm RMS), we observed the saturation of the extracted bunch charge at 1.35 nC (Fig. 3). According to the formula (1) for $\mathbf{E}_{cath}$ ~ 20 /m saturation should start at ~ 0.25 nC per bunch.

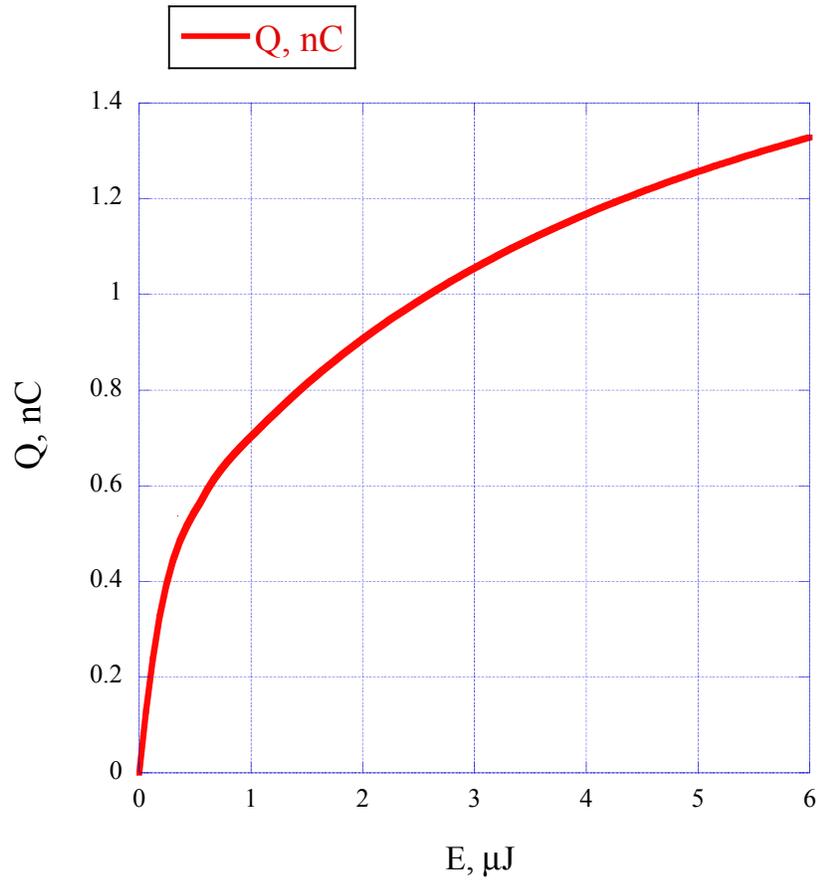

Figure 3. Measured dependence of the generated bunch charge versus laser pulse energy. Measurement was performed at an SRF gun voltage of 1.6 MV.

The measure QE ~0.8% was in good agreement with our expectation, but we observed clear charge saturation at higher energy of the laser pulse. We explain this saturation by part of the bunch profile reaching the critical density defined by accelerated field at the cathode, e.g., eq. (1). Clipping at the charge density defined by $\mathbf{E}_{cath}$=18 MV/m we simulated the dependence of the charge vs. the laser pulse energy, which fits our measured data reasonably well. Increasing the laser spot's size at the cathode to ~2.5 mm increased the maximum generated bunch charge to just above 3 nC. Operating the laser at 5 kHz rep-rate, we generated an average beam current above 15 µA. We also increased the energy of the electron beam to 1.7 MeV (see calibration results below).

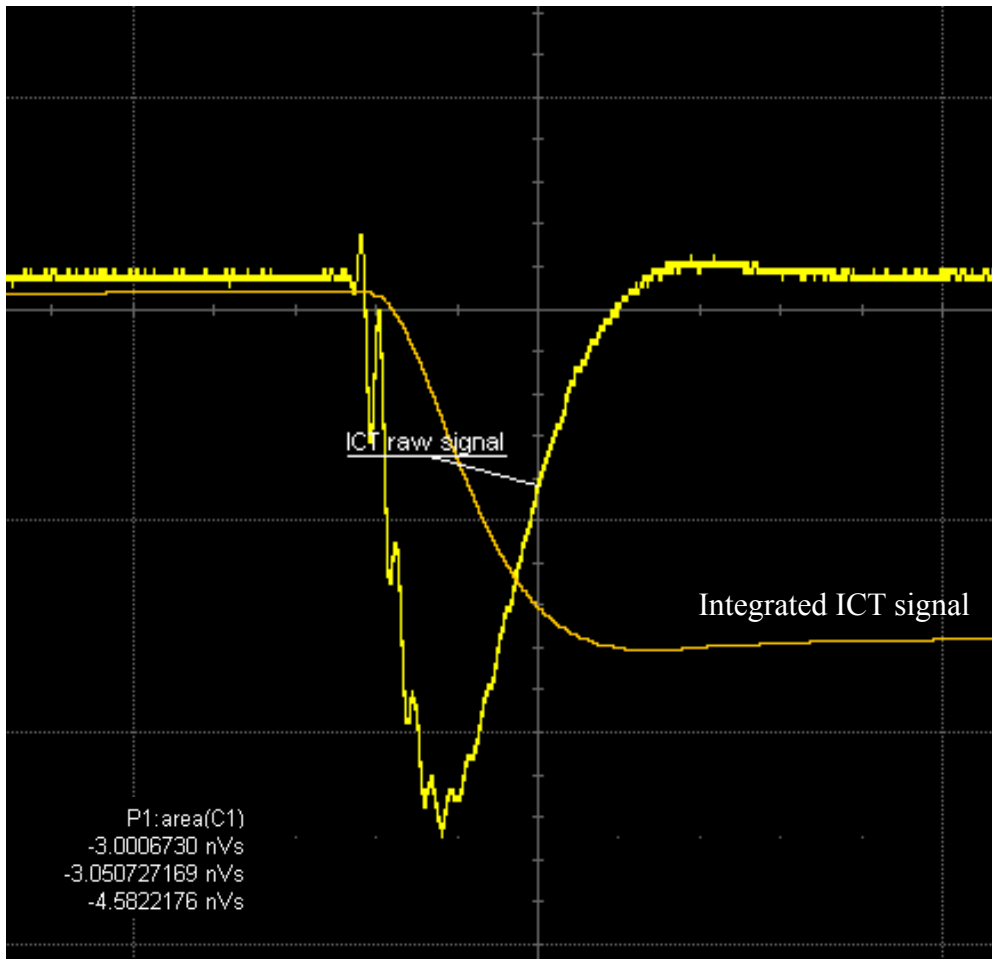

Figure 4. Oscilloscope trace of raw signal from ICT and its integral. The ICT calibration is 0.8 nC per 1 nVs voltage integral [49]. The integrated signal corresponds to a charge of 2.4 nC per bunch.

Our initial attempt to steer the beam to the first YAG screen was not successful – we observed only a sliver of the beam at nearly the maximum settings of our trim dipoles, which indicated an obstacle. Meanwhile, we steered the beam into the wall, which caused a very strong multipacting in the area of our fundamental power coupler at 40 kV RF voltage. Conditioning to the operational voltage caused a lot of vacuum excursions and when the 1.7 MV level was re-established, the cathode QE dropped six-fold. It became apparent that we had poisoned our photo-cathode. Further attempts to propagate the beam to the first viewer screen were unsuccessful. Meanwhile, the cathode's QE deteriorated to about 10 pC per pulse at full laser pulse energy.

With ten hours of operation left before the cryo-system shutdown, we accessed the tunnel to replace the cathode, and to find the obstacle preventing the beam from propagating downstream of the ICT. The Cathode replacement was not successful.

However, we were successful in locating the cause of the beam's obstruction, which was a strong (~ 200 Gs) magnetic field leaking from the vacuum pump's magnet assembly at the laser port. We took the magnets off, and then we were able to propagate the beam (which was only the dark current at that moment) to both YAG screens. We used a vertical trim coil in Trim D to measure the beam's momentum and to calibrate its energy and the SRF gun's accelerating voltage. We used the YAG1 screen (see Fig.1) located 2.77 m downstream of the trim dipole, and turned off Sol2 located between them. Fig.5 shows the results of the measurement. We also had measured the transfer function of the trim dipole to be 1.668

kGs·cm/A (with an accuracy of about 1%). The measured momentum of the beam was 2.02±0.02 MeV/c, corresponding to a kinetic energy of 1.573 MeV (total energy, 2.084 MeV) and the SRF gun's accelerating voltage of 1.612 MV.

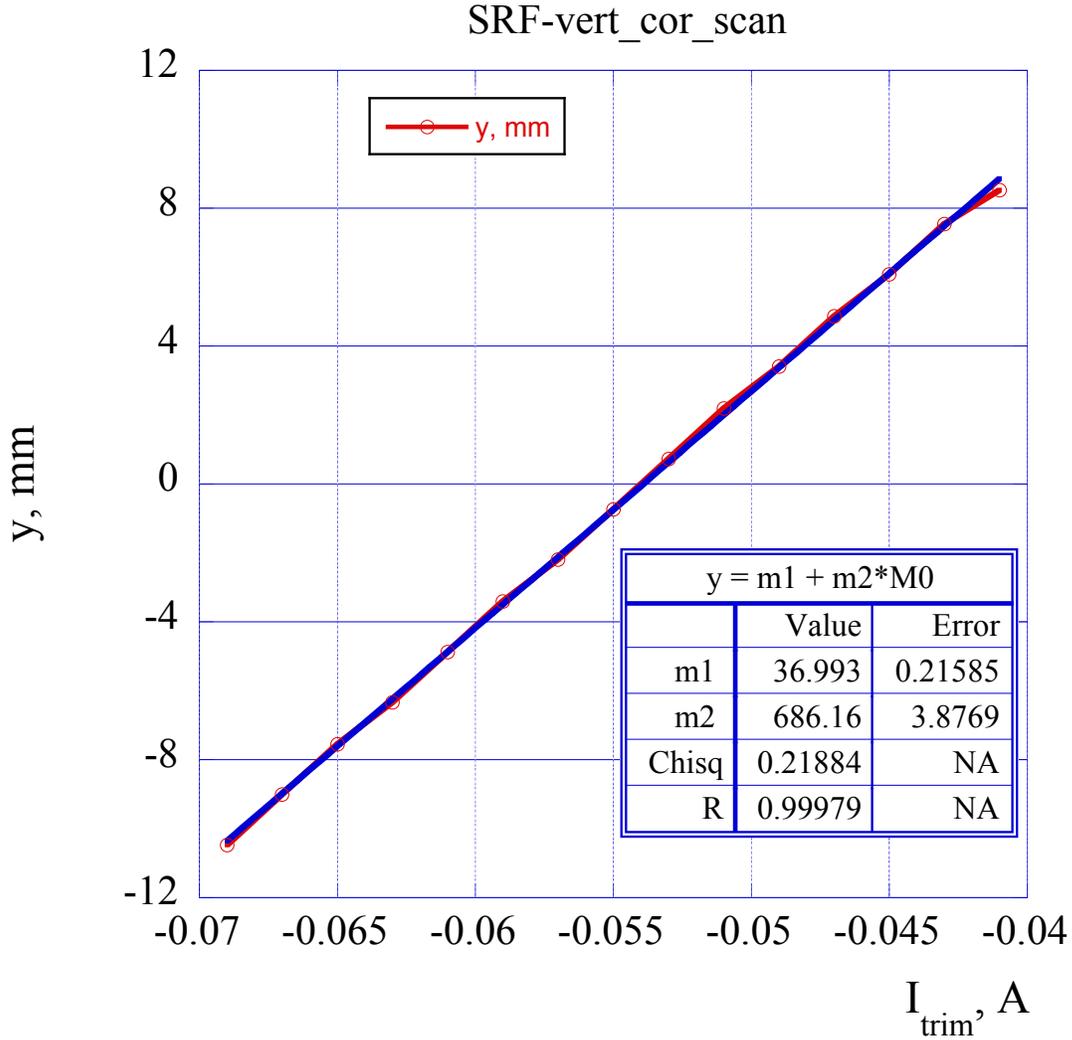

Figure 5. Measured dependence of the vertical beam position at YAG 1 screen as function of the current in trim dipole (Trim D).

Using these measurements as the calibration point, we determined that we generated photo-emitted electron beam with kinetic energies between 1.6 and 1.7 MeV according to our measurements and logged RF pick-up data. We used an expected ratio of 1.02 between the energy of photoelectrons emitted at 78.5° of RF phase, compared with energy of the dark current beam peaking at the crest (i.e., at 90°). We also used this calibration to determine that in the pulsed mode of the SRF gun's operation, the kinetic energy of the beam exceeded 2 MeV. While attempting to increase the energy further, we observed thermal run-off (a precursor of quenching) of the SRF cavity.

While we propagated the beam to our YAG2 beam diagnostic station, we ran out of time to install next photocathode and to measure the photo-emitted beam emittance and energy spread. Since the beam's dynamics in our SRF gun is very close to that in DC guns, we can only suggest that theoretical scaling and experimental observation by the Cornell group [29-31,43] can be applied to the CeC gun. It would mean that according to eq. (2), we should expect a reduction of the transverse emittance by a factor of two by increasing the accelerating field from 5 MV/m to 18 MV/m (for example, reducing normalized transverse emittance to 0.2 mm mrad for 100 pC bunches, if we scale the Cornell's groups experimental results). Our computer simulations support this assumption [56]. Hence, operating our SRF CW gun at 20 MV/m should increase by fourfold the transverse beam brightness.

On the other hand, there is potential increase in the normalized emittance arising from roughness of the photocathode's surface [50]:

$$\varepsilon_{roug} = \sigma_l \sqrt{\frac{\pi^2 a^2 e \mathbf{E}_{cath}}{2mc^2 \lambda_{rough}}} \qquad (3)$$

where $\sigma_l$ is the laser spot's size, $a$ is the amplitude of the surface roughness, and $\lambda_{rough}$ is the characteristic period of the roughness. As previously reported [50-52], this term can be important in high accelerating gradient guns. It also describes a mitigation process available for $CsK_2Sb$ photocathodes by controlling the thickness of the active cathode layers [50,53-54]. We plan to control the cathode surfaces' roughness for attaining ultra-low emittances in the CeC gun. We also plan to measure the emittance of the beam from the CeC gun and fully characterize it as function of the bunch charge when we start operating it again during RHIC Run 16 (after January 2016).

We have proven experimentally that an SRF gun can operate with a high efficiency $CsK_2Sb$ photocathode and generate a CW electron beam with record-high bunch charge accelerated in a record high field at the cathode.

Authors would like to thank the entire team working on coherent electron cooling experiment and especially R. Jecks II, M. DeRosia, T. Lamie, and J. Wyskowski (Niowave Inc.), D. Bevis, K. Brown, L. Hammons, M. Harvey, P. Ingrassia, R. Kellerman, R. Lambiase, K. Mernick, R. Michnoff, A. Pendzick, D. Phillips, P. Sampson, J. Sandberg, T. Seda, J. Walsh, D. Weiss (BNL) We acknowledge financial support for this project from the DOE Office of Nuclear Physics, Facilities and Project Management Division, "Research and Development for Next Generation Nuclear Physics Accelerator Facilities Program" FOA (DE-FOA-0000632) and National Science Foundation (award PHY-1415252).